\documentclass[prb, a4paper, twocolumn, nobibnotes, showpacs, showkeys]{revtex4} %

\usepackage{epsfig}  
\usepackage{graphicx} 
\usepackage{amsfonts}
\usepackage{amssymb}
\usepackage{amsmath}
\usepackage{latexsym} 

\begin{document} 

\author{T.\ H\"{o}rmann and G.\ Brunthaler} %
\email[e-mail: ]{Gerhard.Brunthaler@jku.at} %
\affiliation{Johannes Kepler Universit\"{a}t, Linz, A-4040,
Austria}

\title{Numerical evaluation of the dipole-scattering model for
the metal-insulator transition in gated high mobility Silicon
inversion layers}

\date{\today}

\begin{abstract}
The dipole trap model is able to explain the main properties of
the apparent metal-to-insulator transition in gated high mobility
Si-inversion layers.  Our numerical calculations are compared with
previous analytical ones and the assumptions of the model are
discussed carefully. In general we find a similar behavior but
include further details in the calculation. The calculated strong
density dependence of the resistivity is not yet in full agreement
with the experiment.
\end{abstract}



\maketitle

\section{Introduction} \label{sec:Introduction} 

Since it's discovery in 1995 \cite{Krav94+95}, the metal-insulator
transition in two dimensions (2D) was investigated carefully
\cite{Krav04}, as it's finding is in apparent contradiction to the
scaling theory of localization \cite{Abra79}. According to the
latter, in the limit of zero temperature, a metallic state exists
only in three dimensions, but in two dimensions disorder should
always be strong enough in order to lead to an insulating state
\cite{Abra79}. Nevertheless, high-mobility n-type silicon
inversion layers showed for high electron densities a strong
decrease of resistivity $\rho$ towards temperatures below a few
Kelvin, manifesting the metallic region, and a strong exponential
increase of the $\rho$ for low densities demonstrating the
insulting regime.
A very similar behavior was observed in many other semiconducting
material systems at low temperatures.

After the unexpected finding, several models were suggested in
order to explain the metallic behavior in 2D.  The most important
ones are i) temperature-dependent screening
\cite{SternPRL80,GoldPRB86,DasSarma86}, ii) quantum corrections in
the diffusive regime \cite{Finkel84,Castellani84,Punnoose01}, iii)
quantum corrections in the ballistic regime
\cite{Zala01a,Gornyi04}, and iv) scattering of electrons according
to the dipole trap model \cite{Altsh99PRL}. As there are
argumentations for all that different models in the literature,
we do not want to repeat them here in detail. A clear decision for
one of the suggestions could not been drawn yet
and further work on the models has to be carried out.

The dipole trap model was introduced by Altshuler and Maslov
\cite{Altsh99PRL} (AM) especially for Si-MOS structures, as it is
known that the misfit at the silicon/silicon-oxide interface
produces charged defect states in the thermally grown oxide layer
\cite{Sze81,AFS82,Hori97}. AM could show that a hole trap level at
energy $E_t$ which is either filled or empty, depending on it's
position relative to the Fermi energy $E_F$, can lead to a
critical behavior in electron scattering if $E_t$ and $E_F$ are
degenerate.\cite{FermiEnergy}  This dipole trap model can explain
the main properties of the metal-insulator transitions in gated
Si-MOS structures \cite{Altsh99PRL}.

In this work, we present numerical calculations of the temperature
and density dependent resistivity due to electronic scattering in
the dipole trap model. With these calculations we are able to
checking the analytical
calculations with it's approximations. %
Due to the numerical procedure, we can include further details and
and investigate their influence.

For the analytical calculations AM made a number of assumptions.
These are: a1) the trap states possess a $\delta$-like
distribution in energy, a2) the spatial distribution in the oxide
is homogeneous, a3) the occupied states behave neutral and cause
no scattering of 2D electrons whereas the unoccupied states are
positively charged and lead to scattering (hole trap), a4) a
charged trap state is effectively screened by the 2D electrons so
that the resulting electrostatic potential can be described by the
trap charge and an apparent mirror charge with opposite sign on
the other side of the interface, a5) the scattering efficiency of
the 2D electrons is described by a dipole field of the trap charge
and it's mirror charge, a6) a parabolic saddle point approximation
for the total potential of the trap states was used in order to
perform analytical calculations, a7) the energy of the trap state
$E_{t0}$ is fixed relative to the quantization energy $E_0$ of the
2D ground state inside the inversion potential, and a8) the Fermi
energy $E_F$ in the 2D layer is either independent of or is the
same as in the 3D substrate.

In contrast to AM, our calculations were performed numerically, so
that several limitations of their calculations could be dropped.
Our improvements concern i1) the detailed spacial dependence of
the electrostatic potential is taken into account instead of the
parabolic saddle point approximation, i2) the energy of the trap
state $E_{t0}$ is fixed relative to the conduction band edge
$E_{CB}$. As a result of our calculations, we find a similar
behavior of the calculated resistivity as AM and we calculate in
addition the density dependence of the resistivity.

According to the restricted space in the original AM work, some of
the used equations were not derived there.  We will discuss these
equation and considerations in detail in the main part.  For
better readability of our paper, some details were put into
appendices. Please note that we will use SI units throughout this
work.

\section{Model Considerations and Numerical Calculations} %
\label{sec:Calculations}

\begin{figure}[h]
\begin{center}\leavevmode
\includegraphics[width=0.9\linewidth]{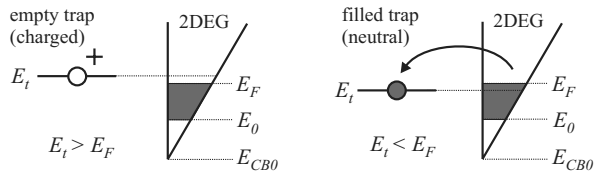}
\caption{Schematic representation of the trap state together with
the 2D electron system (2DES). For $E_t > E_F$ the trap state is
positively charged and scatters electrons in the 2D layer whereas
for $E_t < E_F$ the trap is neutral, i.e.\ it is occupied with an
electron and does not act as a scattering center.  Note that the
main recharging effect comes from the energetic position $E_t$ of
the trap state.  It varies strongly with the applied gate voltage
$V_g$, whereas the Fermi-energy $E_F$ shows only small variations
with changing electron density on the same scale.}
\label{TrapScheme}%
\end{center}
\end{figure}

The misfit at the Si/SiO$_2$ interface layer leads to different
kinds of defects and trap states
\cite{Sze81,AFS82,Hori97,noteTrapStates}.  In the considered AM
model it is assumed that a relative large number of hole trap
states exists.  If such a trap state captures a hole, it is
positively charged, otherwise it is neutral. In Fig.\
\ref{TrapScheme} the trap state is depicted schematically.

As described in the introduction, it is further assumed that a1)
all trap states exist at the same energy $E_{t0}$ if no external
field is applied and posses a2) a spatially homogeneous density
distribution in the oxide layer.  But a potential gradient due to
an applied gate voltage $V_g$ causes a linear increase of the trap
energy position $E_t = E_{t0} + eV_gz/d$, where z is the distance
from the Si/SiO$_2$ interface ($z < 0$) and $d$ is the distance
between gate electrode and that interface (i.e.\ the thickness of
the oxide layer).

\begin{figure}[h]
\begin{center}\leavevmode
\includegraphics[width=0.9\linewidth]{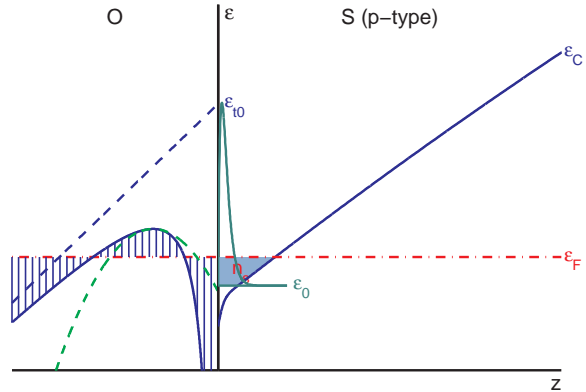}
\caption{Schematic representation of energies in the dipole trap
model.  The dashed blue line on the left represents the trap
energy without, the full blue line with mirror charge potential.
Note that the distance scale on the left and right hand side of
the interface is chosen differently in order to increase
visibility.}
\label{EnergyScheme}%
\end{center}
\end{figure}

For the electrostatic potential inside the oxide layer, also the
screening effects of the inversion layer have to be taken into
account.  For 2D electrons in a Si-(001) layer, the screening
radius
is equal to $a_B/4$.
If the trap distance from the interface $|z|$ exceeds the
screening radius, the in-plane components of the electrostatic
field caused by the charged trap will effectively be screened.  In
that case, the electric field and the potential in the oxide can
be described by the trap charge and an apparent mirror charge with
opposite sign on the other side of the interface (assumption a4).
The potential of the charged state caused by the image charge is
$\Phi = e/(2\cdot4\pi\varepsilon_0\varepsilon_{\textrm{ox}}z)$ in
SI units with $\varepsilon_0 = 8.854 \times 10^{-12}$\,Fm$^{-1}$
and $\varepsilon_{\textrm{ox}} \approx 3.9$, the relative
dielectric constant of the oxide.  Thus the total energy of the
charged trap state can be given as
\begin{equation}
  E_t(z) = E_{t0} + eV_g\frac{z}{d} +
  \frac{e^2}{8\pi\varepsilon_0\varepsilon_{\textrm{ox}}z}
  \hspace{0.8cm} \textrm{for }  z < 0 \textrm{ .}
  \label{eq:Et(z)}
\end{equation}
The last term in Eq.\ \ref{eq:Et(z)} leads to a down bending of
the energetic position towards the interface and causes a maximum
in the total trap energy $E_t(z)$ as shown in Fig.\
\ref{EnergyScheme}.

The trap charge together with it's mirror charge form a dipole
perpendicular to the interface plane. Thus, for distances larger
than $2|z|$, the scattering potential experienced by the 2D
electrons can be described by a dipole field which falls of with
$1/r^3$ (assumption a5).  This is in agreement with the long range
field of a screened Coulomb potential in two dimensions
and leads to a consistent description.
AM have calculated the classical scattering cross section for
momentum relaxation $\sigma_\textrm{m}(E,z)$ for such a dipole
field for electrons with kinetic energy $E$ as
\begin{equation}
  \sigma_\textrm{m}(E,z) =
  2.74(e^2z^2/8\pi\varepsilon_0\varepsilon^*E)^{1/3}
  \label{eq:sig(Ez)}
\end{equation}
with $\varepsilon^* = (\varepsilon_{\textrm{ox}} +
\varepsilon_{\textrm{Si}})/2 \approx 7.9$, the effective
dielectric constant for the 2D electron system (2DES).

Whether a trap state is charged or not, depends on it's
energetical position relative to the Fermi energy $E_F$ (assuming
thermal equilibrium for the occupation). The occupation function
corresponds to a modified Fermi-Dirac distribution, where the
degeneracy of empty and filled states is taken into account. AM
have assumed that the (positively) charged trap state can have
either spin up or down and is thus doubly degenerate, while the
neutral state has no degree of freedom and is not degenerate. From
that the probability of a trap state to be charged follows as
\begin{equation}
  p^+(z) = \frac{1}
  {1 + \frac{1}{2} \exp\left(\frac{E_F-E_t(z)}{kT} \right)}
  \label{eq:p+(z)}
\end{equation}
with $k$ the Boltzmann constant.

For the occupation of the trap states only the relative position
of the trap energy $E_t(z)$ to the Fermi energy $E_F$ is
important.  But the difference $E_F-E_t(z)$ can not be derived
directly -- it has to be calculated from the two individual
energies which depend on different variables.  According to
Eq.~\ref{eq:Et(z)}, the $z$-dependence of the trap energy can be
calculated, but one has to fix it's zero-position $E_{t0}$.  AM
have assumed (a7) a fixed energetical distance of the trap state
relative to the quantization energy $E_0$ of the electronic ground
state in the (nearly triangular) inversion potential.  But $E_0$
depends on the strength and shape of the inversion potential and
via electron-electron interaction on the 2D electron density
$n_s$.
Thus it seems not realistic that the energy of the trap state is
fixed relative to $E_0$, but rather that it is fixed relative to
the energetic position of the conduction band edge $E_{CB}$ (which
is our improvement i2).

Equation \ref{eq:Et(z)} can be used as given, by noting that the
energy is defined relative to the conduction band edge $E_{CB}$.
On the other hand the ground state energy $E_0$ has to be
calculated for the inversion potential, which itself depends on
$n_s$ and the depletion charge $N^-$ and by including the
electron-electron interaction \cite{AFS82}.
As $E_F - E_0$ follows from the electron density $n_s$
\cite{AMnote}, together with $E_0$ one gets the position of $E_F$
relative to the conduction band edge and the difference
$E_F-E_t(z)$ can be used for $p^+(z)$ in Eq.~\ref{eq:p+(z)}.

In the Drude-Boltzmann approximation, the electrical resistivity
$\rho$, equal to the inverse conductivity $\sigma$
\begin{equation}
  \rho = \frac{1}{\sigma} =
  \frac{m^*}{ne^2} \, \frac{1}{\bar{\tau}}
  \label{eq:rhoDrude}
\end{equation}
follows by calculating the effective transport scattering time
$\bar{\tau}$.  The detailed calculation is performed in
Appendix~\ref{Append:Transport}.

As a result one gets that
\begin{equation}
  \frac{1}{\bar{\tau}} = N^+_{\textrm{eff}}\, v(\bar{E})\,
  \sigma_{\textrm{m}}(\bar{E},z_m) \, ,
  \label{eq:tauEff_0}
\end{equation}
can be expressed by the effective values $N^+_{\textrm{eff}}$ for
the number of charged trap states per area, $v(\bar{E})$ the
electron velocity, $\sigma_{\textrm{m}}(\bar{E},z_m)$ the
scattering cross section, and the average electron energy
$\bar{E}$ as given in Appendix~\ref{Append:Transport}.  These
effective values depend on all the important variables of the
systems, i.e.\ on $T$, $V_g$ and so on.  By inserting
Eq.\,\ref{eq:tauEff_0} into Eq.\,\ref{eq:rhoDrude}, one gets
already the dependence of the resistivity $\rho$ on the different
parameters
\begin{equation}
  \rho = \frac{m^*}{ne^2}\, N^+_{\textrm{eff}}\, v(\bar{E})\,
  \sigma_{\textrm{m}}(\bar{E},z_m) \, ,
  \label{eq:rho_0}
\end{equation}
and we have verified equation Eq.\,7 in
Ref.~\onlinecite{Altsh99PRL}.

From here our treatment of the subject is quite different from
that of AM \cite{Altsh99PRL}.  They have evaluated
Eq.\,\ref{eq:rho_0} analytically whereas we perform the
calculation of it numerically. But in order to be able to solve
Eq.\,\ref{eq:rho_0}, AM have used a parabolic (saddle-point)
approximation for the $z$-dependence of the trap energy
(assumption a6). The analytical expression of AM (Eq.\,9) contains
a temperature independent prefactor $\rho_0$ and a temperature
dependent scaling function $R(V_g,T)$.  For their case (A) of the
temperature dependence of $E_F$, they get a critical behavior with
$R(T)$ increasing for $V_g > V_g^c$ and decreasing for $V_g <
V_g^c$ (see Fig.~1 in Ref.~\onlinecite{Altsh99PRL}), similar to
what is observed experimentally. For their case (B), $R(T)$ is
always increasing with temperature and no critical behavior comes
out.

In our numerical evaluation of Eq.\,\ref{eq:rho}, we use the exact
dependence of $E_t(z)$ as given by Eq.\,\ref{eq:Et(z)} (our
improvement i1).  The numerical treatment prevents errors due to
the parabolic approximation, but enables us also to include
further details, which also cannot be solved analytically.

\begin{figure}[h]
\begin{center}\leavevmode
\includegraphics[width=0.9\linewidth]{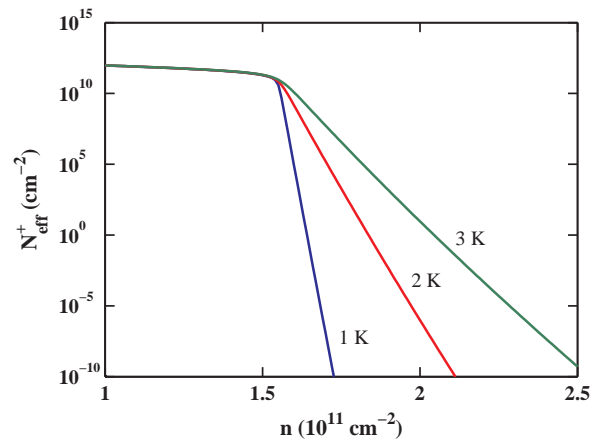}
\caption{Effective number of charged trap states per area
$N^+_{\rm eff}$ vs.\ electrons density (i.e.\ varying gate
voltage) at temperatures of $T = 1$, 3, and 5\,K. %
\label{fig:N+eff(n)} }
\end{center}
\end{figure}

AM have calculated the temperature dependence of the resistivity
in close vicinity of the critical density where the behavior
changes from metallic to insulating behavior.  We have calculated
also the direct density dependence over a larger range for
different temperatures. Fig.\,\ref{fig:N+eff(n)} shows how the
effective number of charge trap states $N^+_{\textrm{eff}}$
depends on $n_s$.

As can be seen there is a very strong variation above $n_s \approx
1.5 \times 10^{11}$\,cm$^{-2}$, where the maximum of the trap
energy $E_t(z)$ is just degenerate with $E_F$.  This strong
variation comes from the fact that as soon as the maximum of
$E_t(z)$ is below $E_F$ only an exponentially small number of
traps is still excited (i.e.\ charged) and the scattering
efficiency decreases accordingly.  As $\rho$ is nearly
proportional to $N^+_{\textrm{eff}}$, such strong variations have
not been observed experimentally.  This discrepancy to the
experiment can possibly be explained that in real 2D Si-MOS
structures either the trap states do not have a $\delta$-like
distribution in energy or that in addition other scattering
sources exist.

\section{Conclusions} \label{sec:Conclusions} %

We have shown that the numerical calculations of the temperature
dependent resistivity give similar results as the analytical
methods by AM.  The strong density dependence of
$N^+_{\textrm{eff}}$ and thus of $\rho$ which follows from the
calculation is not in agreement with experimental findings.  In
order to possibly resolve this discrepancy further calculations
should be performed within the dipole trap model.  The numerical
procedure allows incorporation of further effects and realistic
assumption like energetical broadening of the trap level, special
spatial distributions of the defects, and detailed screening
dependence.

\begin{acknowledgments}
The authors would like to thank A.\ Prinz for his considerations
on the dipole trap model and B.\,L.~Altshuler and D.\,L.~Maslov
for helpful discussions.  The work was supported by the Austrian
Science Fund (FWF) with project P16160 on the ``Metallic State in
2D Semiconductor Structures''.
\end{acknowledgments}

\appendix  

\section{Transport equations} \label{Append:Transport} %

The effective transport scattering time $\bar{\tau}$ in the
Drude-Boltzmann approximation follows from
\cite{AFS82,noteTauWeighting}
\begin{equation}
  \bar{\tau} =
  \frac{\int{d E\, \tau(E) E \, \partial f/\partial E}}
  {\int{d E\, E \, \partial f/\partial E}} \textrm{ ,}
  \label{eq:tau}
\end{equation}
with $\tau(E)$ being the energy dependent scattering time and
$\partial f/\partial E$ the first derivative of the Fermi-Dirac
distribution function $f$.

The transport scattering time $\tau(E)$ has to be calculated by
integration over the individual scattering rates
\begin{equation}
  1/\tau(E) =
  \int_{0}^{d}{ dz\, N_{t3}^+(z)\, v(E)\, \sigma_{\textrm{m}}(E,z)}
  \label{eq:tau(E)}
\end{equation}
with the density of charged traps $N_{t3}^+(z) = N_{t3} P_+(z)$,
$N_{t3}$ the spatially constant density of existing trap states
(both being three dimensional volume densities) and $v(E)$ the
electron velocity.

By inserting the all expressions into Eq.~\ref{eq:tau(E)} one gets
\begin{equation}
  1/\tau(E) = c' N^+_{\textrm{eff}} z_m^{2/3} E^{1/6}
  \label{eq:tau(E)2}
\end{equation}
with the prefactor $c' = 2.74(e^2/8\pi\epsilon_0\epsilon^*)^{1/3}
\sqrt{2/m^*}$, an effective number of positive trap states per
area $N^+_{\rm eff} = N_{t3} \left\langle \Delta z
\right\rangle_{\rm eff}$, the effective width of positive charge
layer $\left\langle \Delta z \right\rangle_{\rm eff} = \int{{\rm
d}z p_+(z) (z/z_m)^{2/3} }$ and the position $z_m =
\sqrt{ed/8\pi\epsilon_0\epsilon_{ox} V_g}$ of the energetical
maximum of the trap energy.

By inserting Eq.~\ref{eq:tau(E)2} into Eq.~\ref{eq:tau}, one gets
\begin{equation}
  \bar{\tau} \propto 1/E_F \int_{0}^{\infty}{ dE\,
  E^{5/6} \partial f/\partial E} . %
  \label{eq:tau(E)3}
\end{equation}

Further an effective energy $\bar{E}$ can be defined so that
formally Eq.\,\ref{eq:tau(E)2} can be preserved for the effective
$\bar{\tau}$, i.e.\ $1/\bar{\tau} = c' N^+_{\textrm{eff}}
z_m^{2/3} \bar{E}^{1/6}$.
A simple calculation gives
\begin{equation}
  \bar{E} = E_F \left[\int_{0}^{\infty} dE\,
  \left( \frac{E}{E_F} \right)^{5/6}\,
  \partial f/ \partial E \right]^{-6} .
  \label{eq:Eeff}
\end{equation}
By further replacing the first derivative of the Fermi-Dirac
function by the identity $\partial f / \partial E = - f (1-f) =
-(4kT \cosh^2((E - E_F)/2kT))^{-1}$ one obtains the same
expression as Eq.\,8 in Ref.~\onlinecite{Altsh99PRL}.

With these relations, the resistivity can exactly be written in
terms of the effective energy $\bar{E}$ as
\begin{equation}
  \rho = \frac{m^*}{ne^2}\, N^+_{\textrm{eff}}\, v(\bar{E})\,
  \sigma_{\textrm{m}}(\bar{E},z_m) \, ,
  \label{eq:rho}
\end{equation}
which corresponds to Eq.\,7 in Ref.~\onlinecite{Altsh99PRL}, but
the individual terms are rewritten according to our definitions
above.  A comparison with Eq.\,\ref{eq:rhoDrude} gives exactly
\begin{equation}
  \frac{1}{\bar{\tau}} = N^+_{\textrm{eff}}\, v(\bar{E})\,
  \sigma_{\textrm{m}}(\bar{E},z_m) \, ,
  \label{eq:tauEff}
\end{equation}
and shows that Eq.\,\ref{eq:tau(E)} can also be rewritten for
effective values.


\begin{thebibliography}{99}
\bibitem{FermiEnergy} Note: In this work we use a temperature %
  dependent Fermie energy $E_F(T)$ instead of the chemical
  potential $\mu(T)$ in contrast to AM, who
  used the nomenclature of $E_F$ beeing identical to the chemical
  potential $\mu$ at  $T = 0$ and using the chemical potential
  $\mu(T)$ for any higher temperature.
\bibitem{noteTrapStates} Note: The type of trap states considered
  by AM might correspond to Si--Si weak bonds, which act like
  donors, i.e.\ being either neutral or positively charged.  If
  the energetic position is deep inside the energy gap, it can also be
  seen as a hole trap state.
\bibitem{AMnote} Note: In this work we calculate the Ferme energy
  inside the 2D layer according to the electron density $n_s$.  AM
  have considered two different cases in their work, the one
  discussed before is their case (B), whereas their case (A)
  assumes that the Fermi energy (chemical potential) of the 2DES
  and the Si substrate coincide and show the same temperature
  dependence.  We do not consider case (A) in detail, as we think
  that case (B) is more realistic.
\bibitem{noteTauWeighting} Note: The weighting of $\tau(E)$ with
  the kinetic electron energy $E$ in Eq.~\ref{eq:tau} fundamentally
  follows from the Drude-Boltzmann approximation \cite{Smith} as
  the Fermi velocity $v_F$ and the shift of the Fermi surface in
  k-space are both proportional to $\sqrt{E}$, which enter into the
  expression for the current $j_x = -e \int{d \vec{k}\,
  n(\vec{k})f(E) v_x }$. The integral in the denominator
  of Eq.~\ref{eq:tau} in 2D is just equal to the Fermi energy $E_F$
  --- also for elevated temperatures.
\end{thebibliography}
\end{document}